\documentclass[manuscript]{emulateapj}  

\usepackage{graphicx}

\def\Ms{{\rm M_\odot}}

\begin{document}

\title{Forming a Primordial Star in a Relic HII Region}
\author{Brian W. O'Shea\altaffilmark{1,2,3}, Tom Abel\altaffilmark{4},
Dan Whalen\altaffilmark{1,2,3} \& Michael L. Norman\altaffilmark{1}}

\altaffiltext{1}{Center for Astrophysics and Space Sciences,
University of California at San Diego, La Jolla, CA 92093, U.S.A.
Email: bwoshea, mnorman, dwhalen@cosmos.ucsd.edu}
\altaffiltext{2}{Department of Physics, University of Illinois in
Urbana-Champaign}
\altaffiltext{3}{Theoretical Astrophysics (T-6), Los Alamos National
Laboratories}
\altaffiltext{4}{Kavli Institute for Particle Astrophysics and Cosmology,
 Stanford University.
Email: tabel@stanford.edu}

\begin{abstract}
There has been considerable theoretical debate over whether photoionization
and supernova feedback from the first Population III stars facilitate or 
suppress the formation of the next generation of stars.  We present results 
from an Eulerian adaptive mesh refinement simulation demonstrating the 
formation of a primordial star within a region ionized by an earlier nearby star.  
Despite the higher temperatures of the ionized gas and its flow out of the dark 
matter potential wells, this second star formed within 23 million years of its 
neighbor's death.  The enhanced electron fraction within the HII region catalyzes 
rapid molecular hydrogen formation that leads to faster cooling in the subsequent 
star forming halos than in the first halos.  This ``second generation'' 
primordial protostar has a much lower accretion rate because, unlike the first 
protostar, it forms in a rotationally supported disk of 
$\sim 10-100~\Ms$.  This is primarily due to the much higher angular momentum of 
the halo in which the second star forms.  In contrast to previously published 
scenarios, such configurations may allow binaries or multiple systems of 
lower mass stars to form.  These first 
high resolution calculations offer insight into the impact of feedback upon  
subsequent populations of stars and clearly demonstrate how primordial chemistry 
promotes the formation of subsequent generations of stars even in the presence of
the entropy injected by the first stars into the IGM.
\end{abstract}


\maketitle

\section{Motivation}\label{motivate}

Calculations performed by Abel, Bryan and Norman (2002; hereafter ABN02)
show that rapid accretion rates driven by molecular hydrogen cooling cause 
the formation of solitary massive protostars in the range of $30$ to $300~\Ms$ 
in minihalos of $10^5-10^6~\Ms$ at redshifts $\ga 20$.  Simulations indicate 
that the hard UV spectra of these 10$^5$ K zero-metallicity stars 
will envelop them in large HII regions several kiloparsecs in diameter (Whalen, 
Abel \& Norman 2004; Kitayama et al. 2004). Over the main sequence lifetime of 
the central star (on the order of 2-6 Myr for the range of
$30-300~\Ms$) half of the baryons within the minihalo are 
driven beyond its virial radius by ionized flows that quickly steepen into shocks.  
These shocks exhibit expansion rates of up to ten times the escape velocity of the 
halo. After the death of the central star, cooling and recombination are out of 
equilibrium in the ionized gas, which results in significant electron fractions
even after its temperature has dropped to 1000 - 2000 K after 20 - 50 Myr. One 
dimensional, nonrotating calculations (Heger et al. 2003) predict two possible 
fates for the primordial stars themselves:  complete destruction by the pair 
instability ($140~\Ms < M_* < 260~\Ms$) which is very energetic and leaves no 
remnant, or direct collapse to black holes above and below this mass range, with 
the added possibility of SN-like precollapse mass ejections by 
pulsational pair instabilities from 100-140 $\Ms$ stars (Heger \& Woosley 2002).

An important question is whether later generations of stars can efficiently form
in the relatively high temperatures and ionization fractions of the relic HII 
regions left by the first stars.  One analytical study (Oh \& Haiman 2003) found 
that the first stars injected sufficient entropy into the early IGM by photoheating 
and supernova explosions to prevent further local star formation in their vicinity.  
Lyman-Werner SUV background radiation is also thought to have contributed negative 
feedback by photodissociating primordial H$_2$ and quenching the molecular hydrogen 
cooling processes allowing the first stars to form (Haiman, Abel \& Rees 2000; Machacek,
Bryan \& Abel 2001).  In this \textit{Letter} we present 
fully resolved simulations that show a second primordial star can form in the relic HII region 
of an earlier Pop III star.  We determine its properties, considering the effect of 
Lyman-Werner radiation from the resultant black hole assuming accretion rates 
consistent with the density fields left by ionized outflows from the parent minihalo.

\section{Simulation Setup}\label{setup}

We carried out simulations using Enzo, a publicly-available Eulerian adaptive 
mesh refinement (AMR) hydrodynamics + N-body code (Bryan \& Norman 1997; O'Shea 
et al. 2004; also see http://cosmos.ucsd.edu/enzo). We initialized a box of 
size $300~h^{-1}$~kpc at $z = 99$ for a cosmology with $(\Omega_M,\ \Omega_
\Lambda,\ \Omega_B,\ h,\ \sigma_8,\ n)=(0.3, 0.7, 0.04, 0.7, 0.9, 1)$.  We 
first ran a simulation with $128^3$ dark matter particles in a $128^3$ root 
grid with 6 total levels of adaptive mesh, refining on a dark matter overdensity 
of 4.0.  This model was run with dark matter alone in order to identify the most 
massive halo that evolves in the simulation volume, which at $z \sim 18$ had a 
mass $\sim 5 \times 10^5 M_{\odot}$.        

We then re-initialized the calculation in the original simulation volume at $z = 
99$ with both baryons and dark matter using a $128^3$ root grid and three 
static nested subgrids, each of which was twice as refined as its parent grid 
and was centered on the Lagrangian volume of the peak that later evolved into 
the identified halo.  The effective root grid resolution was $1024^3$ in this 
volume, which corresponds to a comoving spatial resolution of $\sim 300~h^{-1}$ 
~pc and a dark matter particle mass of $1.8~h^{-1}~M_{\odot}$ in the most highly 
refined region. Every dark matter particle that later enters into dark matter
halos of interest was within this most highly refined grid at the start of the 
simulation.

We started the simulation with this set of initial conditions at $z =
99$ and followed the collapse of the first star, which occurred at a redshift
of 17.76. As a refinement criteria we used a baryon overdensity of 4.0
and a dark matter overdensity of 8.0.  In addition, to ensure appropriate
simulation resolution we mandated that the Jeans length must be resolved by
at least 16 cells at all times, which exceeds the Truelove criterion by a
factor of 4 along each axis (Truelove et al. 1998). At the collapse redshift 
the three dimensional structure was resolved with 8727 grids on nine levels 
containing a total of 49,641,744 unique resolution elements. 

To compute the extent of the HII region of the $120~\Ms$ Pop III star assumed to 
form in the collapse, we interpolated the density, energy, and velocity fields 
from the entire Enzo simulation volume at the formation redshift of this star 
onto a three dimensional grid of fixed resolution with $256^3$ cells for import 
into a static radiative transfer code.  The code utilizes the ionization front 
tracking technique of Abel~(2000) to calculate the boundary of the HII region 
along rays cast outward from the central star by the adaptive ray tracing technique 
of Abel \& Wandelt (2002).  
Within the HII region we set the ionization fraction to unity and the 
H$_2$ and H$^-$ fractions to zero.  We assume that the mean energy of ionization 
for the gas is 2.4 eV, which results in a post-ionization temperature of 
$\sim 18,000$~K when calculated in our multispecies ZEUS simulations.  This is
somewhat cooler than one might expect due to the relatively hard spectrum of
massive primordial stars, and is a result of our use of monochromatic radiative
transfer in the ZEUS code, which underestimates the UV photoheating of the halo
by not taking into account contributions from very high energy photons.  Whalen et
al. (2004) show that an increase in post-front temperatures
results in somewhat higher sound speeds.  These yield higher shock speeds that
promote the photoevaporative flow of gas from the halo in which the first star is 
formed and could in principle affect the dynamics of nearby halos.  We show 
below that in this case the outflow of gas has
a negligible effect on the formation of a second primordial star, which suggests
that our result is at worst only weakly affected by post-front temperature.
Higher post front temperatures will not significantly retard the cooling and 
recombination crucial to the formation of molecular hydrogen.

We approximated the dynamics of the HII region by imposing the one dimensional 
velocity, ionization, density and temperature profiles for a $120~\Ms$ star at
the end of its main sequence lifetime from Whalen et al. (2004) along every line 
of sight from the central star. We modified baryon densities and velocities out to 
$\sim 120$ pc (corresponding to the location of the shock wave in the 1D 
calculation) but changed only ionization fractions and temperatures beyond this
radius out to the boundary of the HII region determined by the ray tracing code.
We then mapped this HII region onto the full hierarchy of grids in the Enzo 
calculation, centering it on the location of the first protostar. This state 
corresponds to only 2.5 million years after the initial star formed (z $\simeq 
17.4$), so we assume that instantaneous ionization is a reasonable approximation
for all gas outside the first halo (which has had the hydro profiles from the 1D
simulations imposed in it).  An important question is whether the satellite halos
are also ionized by the I-front propagating outward from the first star, 
an issue investigated in detail at later redshifts by Shapiro et al. (2004).  
Simulations we performed in 1D in ZEUS-MP indicate that the neighboring halos 
are photoionized by the parent star by the end of its main sequence lifetime.

We then continued the simulation until the collapse of the next protostar, 
which occurs at $z = 16.44$, 22.8 million years later. The final time that we 
analyzed contains 10,710 grid patches on 24 levels with 54,996,560 unique resolution 
elements.  In this calculation we neglect the pulsational pair instability that may eject 
the hydrogen envelope for this star (Heger \& Woosley 2002).  

As a check on our simulation setup we also ran a simulation where we simply
instantaneously ionized the entire simulation volume by raising the baryon temperature
to $\sim 10,000$ K and setting ionization fractions to one and $H_2$ fractions to zero.  This
simulation tests whether the addition of the one dimensional radial profiles from the Whalen 
et al. (2004) calculations changed the properties of the second protostar appreciably.  
We find that the collapse time and accretion rate of the protostar formed in this 
simulation are essentially identical to the results of our full setup, and only discuss
results from the full setup in the rest of this \textit{Letter}.

\section{Results} 
The second primordial protostar forms in a neighboring minihalo approximately 265 
proper parsecs from the location of the halo in which the first star formed (and where the
HII region originated). The halo in which this second protostar forms was completely 
ionized by the first star to a temperature of $\sim 
10^4$ K.  Due to its relatively high density, the center of this halo cools very 
rapidly and molecular hydrogen formation is catalyzed by the extremely high electron 
fraction.  After only a few million years the core of the halo has a molecular 
hydrogen fraction of $\sim 5 \times 10^{-3}$, well above what one would expect for a 
halo which has not been ionized.  This halo is significantly smaller than the first: 
$\sim 2 \times 10^5~\Ms$ rather than $\sim 5 \times 10^5~\Ms$.

\subsection{Comparison of the First and Second Stars}\label{secondprop}
\begin{figure}
\resizebox{3in}{!}{\includegraphics{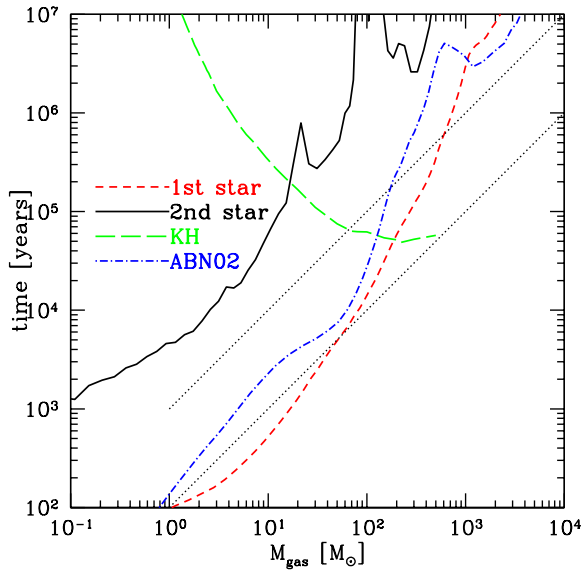}}
\caption{Mass accretion time $t_a = M(r)/\dot{M} \equiv M(r)/(4\pi r^2\rho
v_r)$~as a function of enclosed gas mass. This is at the final output
corresponding to $z=16.437$. The dashed line is the corresponding data dump
of the initial star which had formed at $z=17.67$. The red dashed line
corresponds to the first star to form in this simulation.  The blue
dot-dashed line corresponds to the first star calculated in ABN02.  The
solid black line corresponds to the second star forming in this simulation,
and the green long-dashed line corresponds to the Kelvin-Helmholz time of a
representative star.  The upper and lower black dotted lines correspond to
objects with constant mass accretion rates of $10^{-2}$ and
$10^{-3}~\Ms/$yr, respectively.}
\label{maccrete}
\end{figure}

Figure~\ref{maccrete} compares the mass accretion times of the initial and
second Population III stars formed in this simulation.  In addition, this
figure shows the mass accretion time of the halo in ABN02 and an estimate
of the Kelvin-Helmholz timescale as a function of mass, using values of 
luminosity and effective temperature taken from Schaerer (2002). The upper 
and lower dotted lines correspond to an object with constant accretion rates 
of $10^{-3}$ and $10^{-2}~\Ms/$year, respectively. Our calculation of accretion 
timescales for the initial protostar agrees well with that of ABN02.  The fact 
that the two results are in good agreement even though the ABN02 calculations 
assumed a lower baryon fraction supports the analysis of Ripamonti and Abel (2004) 
showing that all mass scales in these calculations are set by molecular physics. 
Comparison of the accretion rates to the Kelvin-Helmholz timescale provides an 
estimate of $\sim200~\Ms$ for the upper bound of the mass of the star.  The 
accretion timescales suggest a reasonable lower bound of $\sim 80~\Ms$, since this 
much gas will accrete in $10^4$ years, an insufficient time for fusion to begin. In 
contrast, the accretion rate of the second protostar is over an order of magnitude 
lower. This is because the second protostar has a much more pronounced thick disk 
structure than the first protostar. The disk is rotationally supported past a 
radius $\sim 0.01$ pc (corresponding to an enclosed mass of $\sim 10~\Ms$), whereas 
the disk around the first star in the volume is not. Similar accretion timescale 
arguments as before suggest a mass of $\sim 5-20~\Ms$ for the second star, although 
accretion physics will ultimately determine the true mass, particularly given the 
presence of this more pronounced disk.  

Examination of the net angular momentum of the two halos
is illuminating.  The angular momentum of a cosmological halo can be described
by the dimensionless spin parameter: $\lambda \equiv J |E|^{1/2}/G M^{5/2}$
where J is angular momentum, E is the total energy, G is the gravitational constant 
and M is the halo angular momentum.  This is roughly equivalent to the ratio of 
the angular momentum in the halo to the angular momentum needed for the halo
to be completely rotationally supported.  (Padmanabhan 1993)  Typical values of
the spin parameter for cosmological halos are $\sim 0.02 - 0.1$, 
with a mean of $\lambda \simeq 0.05$  
(Barnes \& Efstathiou 1987; Gardner 2001).  We find that the halo
in which the first primordial protostar forms has a spin parameter for the gas and
dark matter of 
$(\lambda_{gas} , \lambda_{dm}) = (0.0275, 0. 0363)$, which is slightly lower than
the mean.  The spin parameter of the second halo is 
$(\lambda_{gas} , \lambda_{dm}) = (0.1079, 0.1607)$, which is atypically high.
Examination of the evolution of angular momentum in the gas of the halos as the 
two protostars form shows that the angular momentum distributions are different
in the two clouds, and if angular momentum is conserved one would expect to see
a centrifugally supported disk that is approximately four times larger in the 
second halo.

\subsection{Black Hole Accretion}\label{bhaccrete}

\begin{figure}
\resizebox{3in}{!}{\includegraphics{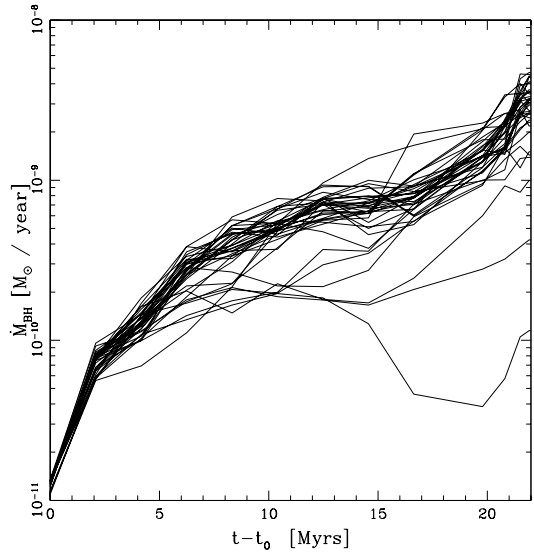}}
\caption{Bondi-Hoyle mass accretion rate around the black hole calculated
from the local gas temperature, density and relative velocity.  Integration
of these curves lead to estimates of growth of the black hole (initially
$120~\Ms$) of that range from 0.009 to 0.032~$\Ms$ over 23 Myrs}
\label{bhacc}
\end{figure}

Here we consider whether accretion onto a relic black hole could generate enough
photodissociative radiation to inhibit H$_2$ formation in the second star's halo.
We assume Bondi-Hoyle accretion (Bondi \& Hoyle 1944) for the 120 $\Ms$ black 
hole that forms after the collapse of the first star to estimate the Lyman-Werner 
flux from its accretion.   This rate depends on the mass of the accretor as 
well as the local gas temperatures, densities, and relative velocities it 
encounters.  To sample the local environment the black hole would traverse over 
the duration of the simulation, we followed the 40 dark matter particles closest 
to the first protostar (within $\sim 0.1$ proper pc) from the end of its main 
sequence lifetime until the collapse of the second protostar.  We tallied the cell
quantities they crossed to compile the accretion rate history each particle would 
have if it were the black hole.  The histories for the 40 black hole proxies appear 
in Figure~\ref{bhacc}. The mass accretion rates grow from $10^{-11}~\Ms$/yr to 
$10^{-8.5}~\Ms$/yr for most of the tracer particles.  

To estimate the effect of Lyman-Werner radiation from the black hole on molecular 
hydrogen formation in nearby halos we assume a canonical 10\% radiative efficiency
for the accretion.  The uppermost accretion curve yields $2.2 \times 10^{37} 
(M/100~\Ms)$ erg/s ($\sim 4500~L_\odot$) for an upper limit to the total luminosity
(which is much lower than the Eddington luminosity of this object, $1.5 \times 
10^{40}$ erg/s, or $\sim 4 \times 10^6~L_\odot$).  Taking this to be a blackbody
spectrum, the flux in the Lyman-Werner band (11.1-13.6 eV) reaching the second 
protostar is $\sim 1.6 \times 10^{-25} (M/100~\Ms)$ erg s$^{-1}$ cm$^{-2}$ Hz$^{-1}$, 
resulting in photodissociation rates that are significantly lower than the formation 
rates of molecular hydrogen there.  The expulsion of gas by ionized flows from the 
first halo prevents higher accretion rates and greater Lyman-Werner fluxes. 
A star in this mass range may shed its envelope just prior to collapse, 
resulting in a smaller black hole and making the results discussed here an upper limit.

\section{Discussion}\label{discuss}

This first high resolution three dimensional simulation of the evolution of gas 
within a primordial HII region demonstrates the crucial role of H$_2$ chemistry 
driven by photoionization in the 
formation of the next generation of stars.  While this has been addressed in 
previous work (Ricotti, Gnedin \& Shull 2002) our simulations are the first with
sufficient resolution to directly examine the formation of individual stars.
Further investigation will be necessary 
to determine if the lower accretion rates leading to the smaller mass of the second 
star are a coincidental feature of this calculation or a general trend of early star 
formation in halos preprocessed by HII regions.  The low accretion 
rate that we observe in this calculation is primarily due to the high initial angular momentum of
the second halo.   

One possible source of error lies in the method and assumptions determining 
whether the neighboring halos are photoionized.  While our 1D results indicate 
that these halos will be ionized, this issue merits further investigation 
with fully 3D simulations.
We further assume that this ionization occurred instantaneously and simply 
ionize the gas outside of the initial halo without changing the total density 
or velocity profiles of nearby halos. Instantaneous ionization appears to be a
 reasonable approximation since the sound crossing time of all of the 
ionized halos is longer than the main-sequence lifetime of the parent star.  
Again, full 3D radiation photo-evaporation simulations will be necessary to determine whether
the hydrodynamic evolution of these halos during the main sequence lifetime of
the parent star is unimportant.

We note that our HII region 
enveloped roughly a dozen minihalos similar to the one that formed the second star.  More
calculations will be required to see if these too form stars.  The evolution of 
the massive disk also merits examination to ascertain whether it breaks up into 
a multiple system or fully accretes to form a single star.  
The situation realized
in our cosmological simulation may lead to objects with initial conditions similar to 
the cases studied by Saigo et al. (2004).
Lower mass second 
generation stars or the possibility of binaries or multiple systems of primordial 
stars would have strong implications for the observability of such objects and
their impact on subsequent structure formation.  Less massive stars might have 
different nucleosynthetic signatures than those of the pair-instability supernovae 
that may occur in the first generation of primordial stars.  The immense size of 
early HII regions could also make the scenario of primordial stars forming in a 
relic HII region much 
more common than extremely massive stars forming in pristine halos.  These two facts
taken together may account for the lack of detection of the characteristic odd-even 
abundance pattern from pair-instability supernovae expected in observations of ultra 
metal poor halo stars (Umeda \& Nomoto 2005 and references therein). How HII regions 
from the first stars may regulate local star formation by suppressing the collapse of 
gas in local halos which have not reached relatively high densities also remains to be 
explored. \\
\\

\acknowledgments{BWO would like to thank Chris Fryer, Alex Heger and Falk
Herwig for useful discussion. This work supported in part by NASA
grant NAG5-12140 and NSF grant AST-0307690 for BWO, DW and MLN. 
BWO has been funded in part
under the auspices of the U.S.\ Dept.\ of Energy, and supported by its
contract W-7405-ENG-36 to Los Alamos National Laboratory.  TA was supported
by NSF CAREER award AST-0239709 from the National Science Foundation.  The
simulations were performed at SDSC and NCSA with computing time provided by 
NRAC allocation MCA98N020.  TA and MLN gratefully acknowledge the Aspen 
Center for Physics for its hospitality during the final phases of this project.
We would like to thank the referee, Nick Gnedin, for suggestions which have
significantly improved the quality of this paper.
}

\bibliographystyle{plainnat}

\end{document}